\documentclass{article}
\usepackage[utf8]{inputenc}
\usepackage{xspace}
\usepackage{hyperref}
\usepackage{soul,color}
\usepackage{amssymb,amsmath,graphicx,verbatim,url,color,rotating,multirow}
\usepackage[margin=1in]{geometry}
\newtheorem{theorem}{Theorem}
\newtheorem{definition}{Definition}
\newtheorem{lemma}{Lemma}

\newcommand{\thesystem}{\textrm{PSI}\xspace}
\begin{document}
\title{Usable Differential Privacy:\\ A Case Study with PSI\footnote{This work is part of the ``Privacy Tools for Sharing Research Data'' project at Harvard, supported by NSF grant CNS-1237235 and a grant from the Sloan Foundation.}}
\author{
Jack Murtagh\thanks{Center for Research on Computation \& Society, John A. Paulson School of Engineering \& Applied Sciences, Harvard University. \texttt{jmurtagh@g.harvard.edu}.}
\and Kathryn Taylor\thanks{\texttt{kathryn.taylor@protonmail.com}} 
\and George Kellaris\thanks{TandemLaunch Inc. \texttt{georgiosk@kalepso.com}}
\and Salil Vadhan\thanks{Center for Research on Computation \& Society, John A. Paulson School of Engineering \& Applied Sciences, Harvard University.  Work done in part while visiting the Shing-Tung Yau Center and the Department of Applied Mathematics at National Chiao-Tung University in Taiwan.  Also supported by a Simons Investigator Award.  \texttt{salil@seas.harvard.edu}.}}

\maketitle

\begin{abstract}
Differential privacy is a promising framework for addressing the privacy concerns in sharing sensitive datasets for others to analyze. However differential privacy is a highly technical area and current deployments often require experts to write code, tune parameters, and optimize the trade-off between the privacy and accuracy of statistical releases. For differential privacy to achieve its potential for wide impact, it is important to design usable systems that enable differential privacy to be used by ordinary data owners and analysts. PSI is a tool that was designed for this purpose, allowing researchers to release useful differentially private statistical information about their datasets without being experts in computer science, statistics, or privacy.  We conducted a thorough usability study of PSI to test whether it accomplishes its goal of usability by non-experts. The usability test illuminated which features of PSI are most user-friendly and prompted us to improve aspects of the tool that caused confusion. The test also highlighted some general principles and lessons for designing usable systems for differential privacy, which we discuss in depth.
\end{abstract}

\section{Introduction}
In recent years both science and industry have seen an unprecedented explosion in data collection and analysis. Much of these data contain sensitive information about humans, whether it be medical records, cell phone usage, or research datasets. As these data are shared and mined, there is an urgent need for mechanisms that protect the privacy of individuals. Differential privacy \cite{dwork06,DworkKeMcMiNa06} is an attractive approach to preserving privacy while maintaining utility for statistical analysis. By injecting a carefully calibrated amount of random noise into statistical calculations, differentially private algorithms guarantee that they do not leak individual-level information even to adversary's with arbitrary auxiliary information about the data subjects. Since its inception over a decade ago, differential privacy has developed into a rich theory showing that many statistical data analysis tasks are compatible with its strong privacy protections.

There are a handful of deployments of differential privacy in practice, discussed in Section \ref{sect:systems}, including Apple, Google, and the U.S. Census \cite{apple,erlingsson14,mac08}. In all of these cases, teams of technical experts customized and optimized differentially private algorithms for their specialized use cases. Some systems aim for wider use by developing programming platforms to help people write differentially private code \cite{mcsherry09,roy10,mohan12}. Both types of applications suffer from the need for technical expertise to enjoy the benefits of differential privacy, a complex mathematical subject that can require experts to write code, tune algorithms, set privacy loss parameters and manage a privacy budget (both defined in Section \ref{sect:dp}), and interpret the tradeoff between privacy and accuracy. The problem is that such expertise is rare, while privacy-sensitive data releases are not. There simply are not enough differential privacy experts to play a significant role in every instance of data sharing. 

PSI \cite{PSI} is a system that was designed to address this problem by making differential privacy usable by non-experts.\footnote{A prototype of PSI can be found at \href{http://beta.dataverse.org/custom/DifferentialPrivacyPrototype/}{http://beta.dataverse.org/custom/DifferentialPrivacyPrototype/}.} It allows data owners to generate and publish statistics describing their datasets with the strong protections of differential privacy. Using a graphical user interface, data owners choose a privacy level for their dataset and then select what they want to release from a variety of built in statistics. The interface elicits information needed for the differentially private algorithms and gives users a sense of how accurate they can expect their outputs to be. Wherever possible, the system translates technical concepts from differential privacy (e.g. the privacy loss parameters $(\epsilon,\delta)$) into more interpretable terms (e.g. accuracy as measured by $95\%$ confidence intervals) so that lay users can make responsible and accurate choices without being experts in computer science, statistics, or privacy.  

We conducted a usability study of \thesystem to determine whether it achieves its goal in making differential privacy usable by non-experts. Our study involved a total of 28 participants across three study phases, and yielded significant insights for the improvement of the tool and, we believe, for efforts to bring differential privacy into practice more generally. Performance in the usability study was encouraging overall. Most participants were able to complete the typical workflow of a \thesystem user by setting privacy loss parameters, selecting statistics, allocating a privacy budget, and using other advanced features. Participants gave \thesystem a strong score on a widely used questionnaire assessing system usability (see Section \ref{sect:sus}). The user tests also highlighted directions for improving \thesystem by showing us common mistakes and sources of confusion. In response to these observations, we improved the documentation of the tool and implemented a tutorial mode, tidied the interface, improved error-checking on user inputs, made help buttons easier to find, and made a host of cosmetic adjustments.

Tracking common mistakes and successes in the user study also taught us general lessons about bringing differential privacy to practice beyond \thesystem. We discovered that users of a system like \thesystem are highly inclined to follow suggestions provided by it such as guidelines for setting privacy loss parameters, and consequently that these kinds of recommendations should be made with extreme care. We learned that introducing a concept as technical as differential privacy in a user interface can lead to some confusion, but a well-designed tool can help even the most confused users achieve their ultimate goals of sharing useful statistics while protecting privacy.

Most encouragingly, when we asked participants how relevant they think a tool such as \thesystem is for researchers working with sensitive datasets on a scale from 1 to 5, the average response was 4.3. We believe this reflects that privacy is not just important to us, but also to practitioners who shoulder the responsibility of protecting the privacy of research subjects. Hence, there is a pressing need for a practical and usable tool for privacy-protective data sharing. 

We hope that these lessons will inform future attempts to deploy differential privacy in various user-facing contexts. Before presenting our insights, we explain the concept of differential privacy in Section 2.1, discuss examples of other differentially private systems in Section 2.2, and then introduce \thesystem in detail in Section 3. Section 4 describes our usability test. The heart of the paper’s contribution lies in Section 5, where we discuss in detail the relevant lessons learned from our usability study of a differentially private system. The last section presents some open questions and offers a few final thoughts.

\section{Background}
We begin with a description of differential privacy and a discussion of its deployments in actual systems. 

\subsection{Differential Privacy}
\label{sect:dp}
Suppose a researcher gathers sensitive information about people and stores it in a dataset $D$. Her goal is to publish statistics computed from $D$, without violating the privacy of the respondents. The idea behind differential privacy is to carefully introduce noise into the statistical calculations so as to obscure the presence of any individual in $D$ while maintaining accuracy in the published statistics i.e., they do not differ greatly from the actual statistics. 

Think of a dataset $D$ as a table where the rows correspond to individuals and the columns correspond to variables or attributes about the individuals. We call two datasets $D, D'$ \emph{neighboring} if they differ on only one row. Differential privacy is a property of a randomized algorithm $\mathcal{M}$ that takes datasets as input and outputs noisy statistics in some output space $\mathcal{O}$. A data owner runs $\mathcal{M}$ on her gathered data $D$, and publishes $M(D)=o \in \mathcal{O}$. Differential privacy is defined as follows.

\begin{definition}[Differential privacy~\cite{dwork06,DworkKeMcMiNa06}]\label{def:epsilon_DP}  
A randomized algorithm $\mathcal{M}$ satisfies $\boldsymbol{(\epsilon,\delta)}$\textbf{\emph{-differential privacy}}, where $\epsilon,\delta \geq 0$, if for all $O \subseteq \mathrm{Range}(M)$, and all neighboring datasets $D,D'$, it holds that
$$
	\mathrm{Pr}[\mathcal{M}(D)\in O] \leq e^\epsilon \cdot \mathrm{Pr}[\mathcal{M}(D') \in O] +\delta
$$
where the probabilities are over the randomness of the algorithm $\mathcal{M}$.
\end{definition}  

In other words, an algorithm $\mathcal{M}$ is differentially private if the probability that $\mathcal{M}$ produces an output in a set $O$ when a particular person is in a dataset $D$ is roughly the same as in the case where this person's data is removed from $D$, resulting in a neighboring dataset $D'$. Individual privacy is ensured by the fact that an adversary cannot distinguish statistics computed on $D$ from those computed on $D'$. 

$\epsilon$ and $\delta$ are called the \emph{privacy loss parameters}. The smaller their values, the more noise is introduced to the statistical outputs, and the stronger the privacy guarantees. This \emph{privacy/accuracy trade-off} is inherent in differential privacy and much of the effort in designing differentially private algorithms goes toward maximizing the accuracy of outputs for each desired level of privacy. In practice $\epsilon$ should be set to a small constant (e.g. $\epsilon=.5$). $\delta$ can be interpreted as the probability of a blatant privacy violation and so should be set quite small (e.g. $\delta=10^{-6}$).

Typically in practice, one wants to release more than one statistic about a dataset. An important feature of differential privacy is that the releases from multiple differentially private algorithms still maintain privacy. This phenomenon is captured in `composition theorems' like below.

\begin{theorem}[Composition~\cite{dwork06,DworkKeMcMiNa06}]\label{theo:comp}
Let $\mathcal{M}_1, \ldots, \mathcal{M}_k$ be algorithms, where $\mathcal{M}_i$ satisfies $(\epsilon_i,\delta_i)$-differential privacy for each $i\in\{1,\ldots,k\}$. Let $\mathcal{M}$ be another algorithm that executes $\mathcal{M}_1(D), \ldots,$ $\mathcal{M}_k(D)$ on a dataset $D$ using independent randomness for each $\mathcal{M}_i$, and returns the set of outputs of these algorithms. Then, $\mathcal{M}$ satisfies
$\left(\sum_{i=1}^k{\epsilon_i}, \sum_{i=1}^k{\delta_i}\right)$-differential privacy.
\end{theorem}

The above theorem can be improved to give even tighter privacy guarantees for any set of differentially private algorithms \cite{DRV10,MurtaghV16}. Moreover, it allows us to view $(\epsilon, \delta)$ as a \emph{privacy budget} that is distributed among the $k$ algorithms. In many cases it makes sense to set a privacy budget for a dataset based on the sensitivity of its contents and then release as many accurate differentially private statistics about the data as possible within the budget. We sometimes refer to the privacy budget $(\epsilon,\delta)$ as \emph{global privacy loss parameters} to distinguish it from the \emph{individual privacy loss parameters} $(\epsilon_1,\delta_1)\ldots(\epsilon_k,\delta_k)$ assigned to each statistic as in Theorem \ref{theo:comp}.

We end this section with another useful fact in the theory of differential privacy that is discussed more in Section \ref{sect:secofsamp}. Suppose we draw a random sample of rows from a database, and publish statistics about the sampled data instead of the full dataset. Sampling increases the uncertainty of the adversary about a user being in the sample. As a result, less noise is needed in order to satisfy the same level of privacy protection for the sampled data. 

\begin{lemma}[Secrecy of the sample~\cite{KLNRS08, AdamSecrecyPost}]
\label{lem:secofsamp}
Let $M$ be an $(\epsilon,\delta)$-differentially private algorithm for datasets of size $n$. Let $M'$ be a randomized algorithm that takes as input a dataset $D$ of size $m\geq n$, and then runs $M$ on a dataset $D'$ obtained by selecting a uniformly random subset of $D$'s records of size $n$. Then, $M'$ is 
$((e^\epsilon-1)\cdot (n/m),\delta\cdot (n/m))$-differentially private.
\end{lemma}

For small $\epsilon$, $e^{\epsilon}-1\approx \epsilon$, so the lemma says that subsampling improves the privacy loss parameters by a factor of roughly $m/n$. Consequently, if we know that we are given a random sample of a large dataset or population, then we can achieve the same privacy level by injecting less noise into the output. We describe a practical use of Lemma \ref{lem:secofsamp} in Section \ref{sect:secofsamp}.

\subsection{Differentially Private Systems}
\label{sect:systems}
There are several examples of deployed systems that use differential privacy. We divide them into two broad categories: i)~algorithms that perform specific statistical tasks using data that are pre-defined by a third party expert, and ii)~programming languages, where the programmer can choose the privacy loss parameters, tasks, and data themselves. In the latter category, only users with technical and programming backgrounds can produce differentially private outputs.

The most well-known systems that report statistics on pre-defined data using differential privacy are the U.S. Census OnTheMap, Google's RAPPOR for telemetry, Google's sharing of historical traffic statistics, and Apple's MacOS/iOS for its intelligent assistance and suggestions technology. 

Specifically, the U.S. Census OnTheMap \cite{mac08} reports differentially private statistics on the source-destination pairs of commuters in the U.S. Google's RAPPOR \cite{erlingsson14} is used for the collection of statistics (e.g. homepage settings, running processes, etc.) from a large number of users of the Chrome web browser. Moreover, Google uses differential privacy for sharing historical traffic statistics\footnote{https://europe.googleblog.com/2015/11/tackling-urban-mobility-with-technology.html} such as average traffic speed, relative traffic volumes, and traffic trajectory patterns. Apple's\footnote{https://images.apple.com/au/privacy/docs/Differential\_
Privacy\_Overview.pdf} differential privacy is employed on iOS 10 for collecting user data in order to improve QuickType and emoji suggestions, Spotlight deep link suggestions, Lookup Hints in Notes, and on MacOS Sierra to improve autocorrect suggestions and Lookup Hints. Finally, Uber\footnote{https://medium.com/uber-security-privacy/differential-privacy-open-source-7892c82c42b6} utilizes differential privacy for internal data analytics in order to execute private SQL aggregate queries (e.g., counts, sums, and averages). The source code of the latter is available online.\footnote{https://www.github.com/uber/sql-differential-privacy}

The above systems are all designed for very specific data sources and analytic purposes, and do not provide a general-purpose tool that allows lay users to make effective use of differential privacy. On the other hand, there are fully parameterized programming languages that allow a programmer to produce differentially private outputs on sensitive inputs. The most well-known solutions are PINQ, Airavat, Fuzz, and GUPT.

PINQ \cite{mcsherry09} is a programming language that guarantees differential privacy as long as the input data are
accessed exclusively through specific PINQ queries. Airavat \cite{roy10} is a Hadoop-based MapReduce programming platform whose primary goal is to offer end-to-end security guarantees, while requiring minimal changes to the programming model or execution environment of big data computations. Fuzz was introduced by Haeberlen et al. \cite{haeberlen11} to alleviate a specific type of attack that on PINQ and Airavat. Fuzz has functionality similar to that of PINQ, but it avoids using primitives susceptible to known attacks and implements a query processor that ensures that each query does not reveal sensitive information even implicitly. Finally, GUPT \cite{mohan12} aims at increasing the accuracy of the periodic publications of differentially private statistics by allowing the privacy of older statistics to degrade through time, and thus enabling a more efficient allocation of the privacy budget among different applications. These systems are designed for users who have both programming and statistical skills, and also provide little assistance with setting privacy loss parameters or privacy budgeting. 

The details described above illustrate that differential privacy has been used for a range of applications, but that these systems have not been designed to be accessible or navigable by inexperienced or non-technical users. There has, until the creation of PSI, been a gap between the many potential applications of differential privacy and the existence of widely usable systems.

\section{PSI}
In contrast to previous systems, \thesystem \cite{PSI} is designed for users with limited background in computer science or privacy who wish to share or explore sensitive datasets. There are two main actors who may use \thesystem:
\begin{enumerate}
\item \textbf{Data owners} are people who hold a privacy-sensitive dataset and are looking to share differentially private statistics about their data with others. Data owners use an interface called the \emph{Budgeter} to set privacy loss parameters for their dataset, select statistics to release, and allocate their privacy budget to get a favorable privacy/accuracy tradeoff. Data owners have the option to reserve a portion of their privacy budget for future users called data analysts. 
\item \textbf{Data analysts} are people who want to explore a privacy-sensitive dataset but do not have access to the raw data. They use the \emph{Interactive Query Interface} to view the statistics released by the data owner and, if they were given a portion of the budget for the dataset, can run their own queries on the dataset.
\end{enumerate}

Our usability test only evaluated the Budgeter interface and the behavior of data owners. The Interactive Query Interface is still under development and will need independent testing. \thesystem is a tool that was developed to reconcile the growing tension between the open data movement in the social sciences and concerns over the privacy of individuals. Sharing of research data has become a common and sometimes even expected practice in science so that results can be validated, replicated, and extended, but it carries risks of inadvertently leaking sensitive information about research participants. 

\thesystem helps data owners release privacy preserving statistics describing their dataset by distributing a privacy budget across different possible statistical calculations with the strong guarantees of differential privacy. The main goal of \thesystem is to make differential privacy usable by non-experts so that data owners can share their data \emph{and} protect privacy. 

\begin{figure*}[htb]
 % \begin{minipage}[c]{0.76\textwidth}
    \includegraphics[width=\textwidth]{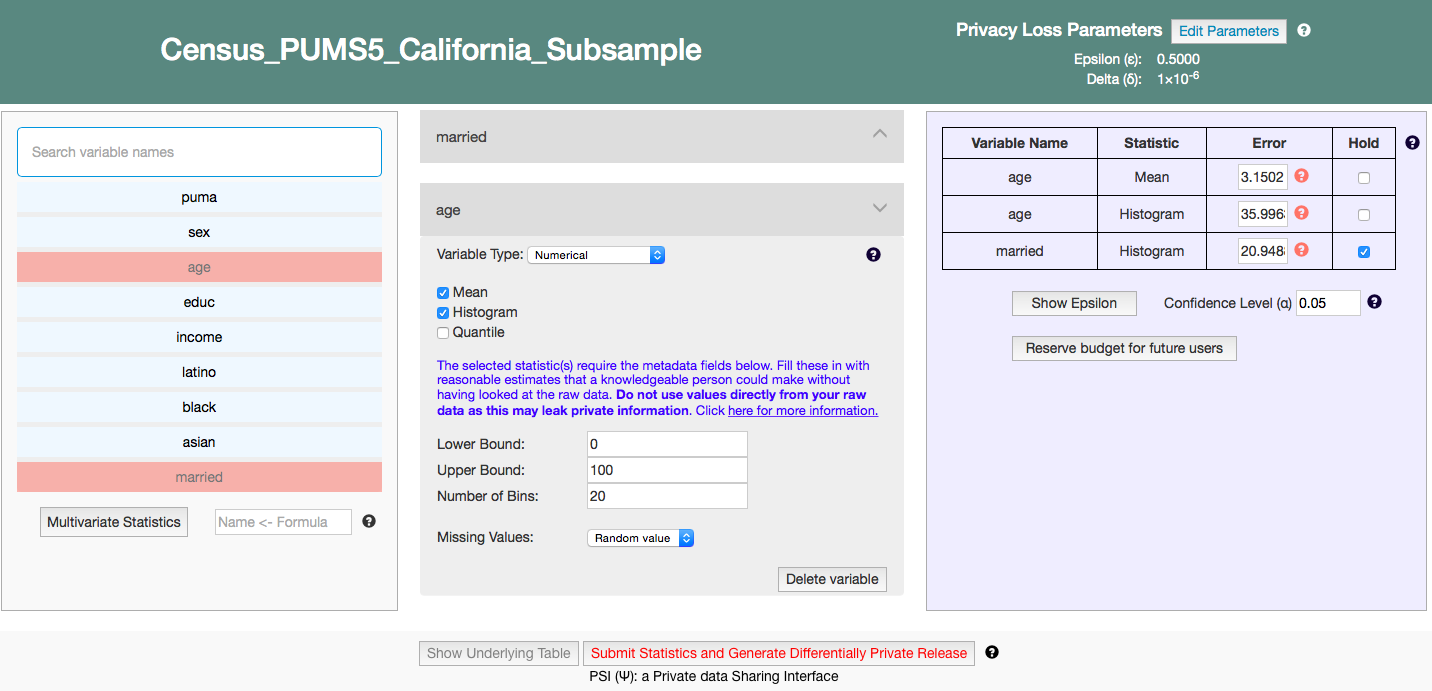}
 % \end{minipage}\hfill
 % \begin{minipage}[c]{0.23\textwidth}
    \caption{The PSI user interface. Variables in the dataset are listed in the left panel. Selecting one spawns a bubble in the middle panel where users select statistics and supply metadata. Part of the privacy budget is automatically assigned to new statistics and worst case error estimates are displayed in the table on the right. 
    } \label{fig:interface}
\end{figure*}

\subsection{Overview}
\label{sect:psioverview}
To begin using \thesystem, a data owner uploads a dataset into the system and chooses the values of the privacy loss parameters $\epsilon$ and $\delta$ for her dataset. The tool provides an introduction to differential privacy and recommendations for parameters in accessible but accurate documentation.
At this same time, the user can opt to use a secrecy of the sample feature. As described in Lemma \ref{lem:secofsamp} if the data are a random and secret sample from a larger population of known approximate size, the user can get boosts in accuracy for free. The data owner can enjoy this benefit if her data satisfy the requirements and she provides an approximation of the population size.

Figure \ref{fig:interface} shows a screenshot of the interface with the Public Use Microdata Sample (PUMS) California Dataset\footnote{https://www.census.gov/census2000/PUMS5.html} loaded. After choosing parameters, the user selects statistics to release one by one. She starts by selecting the variable on which to compute a new statistic in the left panel. This spawns a box in the center panel for that variable. Then, she fills in the variable type (e.g. boolean, numerical, categorical) and selects a statistic from the supported options. Currently, these include univariate summary statistics (e.g. mean, histogram, quantile, CDFs), low dimensional regressions (including linear, logit, probit, and poisson), and difference-of-means tests for causal inference. 

Many differentially private algorithms require auxiliary information about the variables in order to run. We call this \emph{metadata}. For example, in order to release a differentially private mean, the algorithm needs to know an upper and lower bound on the data in order to calibrate its addition of noise. It is important that such metadata be supplied in a data-independent way. For example, a reasonable a priori range estimate for any age variable would be 0 to 150. Using empirical minimum and maximum ages in the specific dataset could violate differential privacy (for example, if the maximum is reported as 115, then one could deduce that Kane Tanaka of Japan, the oldest living person\footnote{https://www.newsweek.com/who-worlds-oldest-person-miyako-chiyo-dies-117-passing-title-kane-tanaka-1044901}, is in the dataset. This could be sensitive, for example if the dataset consisted of people diagnosed with a particular disease.) So once a user selects a statistic, she may be asked to supply certain pieces of metadata depending on the needs of the algorithm for that statistic. The user should pick metadata that is as accurate as possible while remaining data-independent. For example, if a supplied range of values is smaller than the true range in the data, the data is clipped to conform to the supplied range, which could lead to inaccurate statistics. If the supplied range is too large, more noise will be added to protect privacy, also leading to less accurate outputs. 

As statistics are selected, the privacy budget gets automatically divided across them. Rather than show individual privacy loss parameters for each statistic, which may be hard to interpret for lay users, \thesystem displays associated error estimates for statistics, a user-friendly idea that was introduced in GUPT \cite{mohan12}. The estimates are displayed in the table on the right panel to provide a sense of the privacy/accuracy trade-off for the current selections. There is also a confidence level, $\alpha$, representing an upper bound on the probability that the actual error in the final computed statistics exceeds the error bound listed in the table. The interpretation of these accuracies differs across statistics. For example, for histograms, the error reports an upper bound on the difference between the noisy count and the true count for each bin that will hold with probability at least $1-\alpha$. The user can change the confidence level $\alpha$ in the right panel, under the error estimates. The errors reported to users are a priori estimates of worst case error and do not depend on the dataset. The interface only accesses the raw data after the session is complete and the user's budgeting selections are finalized. 

The tool defaults to distributing the privacy budget evenly across the chosen statistics. Because of the privacy/accuracy trade-off, however, spending a larger percentage of the privacy budget on a statistic will improve its accuracy. Not all differentially private algorithms strike the same balance between privacy and accuracy. Some statistical information about a dataset may be more important than others, so an even distribution of the privacy budget is not the optimal allocation for every dataset and data release. The user has the option to redistribute her privacy budget by directly editing the reported error estimates in the right panel. When the user reduces the desired error for a statistic, the tool automatically determines what portion of the privacy budget would be required to achieve that error and accordingly assigns more of the budget to the chosen statistic. In order to stay within the fixed privacy budget, the tool then reduces the amount being spent on every other statistic to accommodate the improved error on the chosen one. This action causes all of the error estimates to update in real time so that the user can see the consequences of her request for more accuracy. 

If the user wishes to keep an error level fixed for a particular statistic, she can check the ``hold'' box next to this statistic in the right panel. Further redistributions of the privacy budget will not affect held statistics. This budgeting process continues until the user is satisfied with her chosen statistics and the associated error estimates, at which point she submits the page and the actual differentially private algorithms are run on the raw data according to the parameters she set. 

In the right panel of the interface there is the option for reserving budget for future analysts. Using a slider, the data owner can choose a portion of the global privacy loss parameters that should not to be used when calculating the statistics. This budget can be spent at a later stage by other data analysts in order to compute additional statistics if needed. Other options available to users include the ability to change the privacy budget mid-session, delete statistics that have already been selected, and bring up documentation by clicking any of the question mark buttons around the screen.

\subsection{Goals}

The goals of \thesystem, as described in \cite{PSI}, are threefold: i)~accessibility by non-experts, ii)~generality, and iii)~workflow-compatibility. These goals are described in detail below.

\textit{Accessibility by non-experts}: The system allows data owners that are not experts in privacy, computer science, or statistics, to share their data with others without violating the privacy of their research subjects. The target audience consists of social scientists that wish to allow other analysts to explore their gathered data and run replication studies, but are cognizant of confidentiality. PSI offers them this functionality without requiring them to learn concepts from the field of computer science and ensures them that the privacy of the examined population is preserved.

\textit{Generality}: \thesystem is dataset-independent. It is designed to support general queries that apply in a wide range of data sources or domains. At the same time, these queries are the most widely used in the social sciences and allow for identifying the basic characteristics of each dataset. Its set of supported queries  are integrated with existing statistical software designed for social science researchers, namely the TwoRavens \cite{honaker2014statistical} graphical data exploration interface. 

\textit{Workflow-compatibility}: The system is part of the workflow of researchers in the social sciences, while offering access to sensitive data. It is designed to be integrated with existing and widely used data repository infrastructures, such as the Dataverse project \cite{Crosas11, King07}. It is also part of a broader collection of mechanisms for the handling of privacy-sensitive data, including an approval process for accessing raw data, access control, and secure storage. 

The usability test described in the remaining sections of this paper assesses whether \thesystem achieves its first goal of usability by non-experts. Further discussion of the other goals can be found in \cite{PSI}.

\section{Usability test}
This section describes the procedures and results of our usability tests of \thesystem. We saw 28 participants in total over the course of three study phases: a pre-pilot phase ($n=3$), a pilot phase ($n=5$), and a full study ($n=20$). Improvements to the tool and to the study protocols were made after each round in response to participants' feedback and performance. This section focuses on the full study because only those participants used the latest version of the software. 

We recruited participants by sending calls out to social science listservs and posting on student job boards at local universities. All participants were adults (over 18 years old) with data analysis experience either through college courses, professional work, or research. Participants received $\$20$ gift cards to Amazon for about an hour of their time. The study was approved by a university Institutional Review Board. The education level and degree of familiarity with differential privacy varied within the group and are summarized in Table \ref{tab:demographics}.
\begin{table}
\begin{center}
\begin{tabular}{|r|c|}
\hline
 \textbf{Education level~~~~~~~} & Number of participants \\
\hline
Some college & 2  \\
\hline
Bachelor's degree & 5 \\
\hline
Master's degree & 10 \\
\hline
PhD &  3\\
\hline
\textbf{Familiarity with DP~} &   \\
\hline
Unfamiliar &  7 \\
\hline
Somewhat familiar &  10 \\
\hline
Familiar & 2 \\
\hline
Very familiar &  1  \\
\hline
Expert &  0  \\
\hline
\end{tabular}
\end{center}
\caption{Education level and familiarity with differential privacy in the full study sample.}
\label{tab:demographics}
\end{table}%

\subsection{Study Protocol}
\label{sect:protocol}
After the consent process and a brief background questionnaire, participants were presented with a laptop with \thesystem displaying its introductory text. The text was written to describe the purpose of the tool and key concepts that are important for using the tool (e.g. privacy/accuracy trade-off, composition, and secrecy of the sample) in a digestible yet accurate way. On average, participants took 4 minutes and 26 seconds to complete the reading. 

Next, participants were given a scenario (in written and verbal form) intended to simulate the mindset of a typical \thesystem user (a data owner) trying to release differentially private statistics on a dataset. They were given a toy dataset in Excel with 1000 records containing variables for age, sex, income, education level, race, and marital status and were told that it was a random sample from a county in California with population 700,000. The goal of the scenario was to advertise this dataset to other social scientists who study the relationship between race and income in different age groups. Participants were given time to browse the dataset and familiarize themselves with its contents.

After reviewing the dataset, we turned on screen and audio recording and instructed participants to speak their thoughts aloud as much as possible for the remainder of the user test. They were encouraged to voice their intentions, confusions, and questions as they navigated the tool although we would not always answer the questions if doing so would affect the validity of the results (e.g. revealing the solution to a task.) The audio recordings provided valuable qualitative feedback about participants' experience with the tool. The screen recordings also allowed us to confirm the notes we took about participants' performance during each session.

Once recording was on, participants were asked to set the privacy loss parameters for the given scenario. We did not specify the choices but participants had access to PSI's introductory text, which contained some parameter recommendations. At this time the interface also gave participants the option to fill in an approximate population size if they wished to use the secrecy of the sample feature.

The next step, and the heart of the study, was a sequence of 11 tasks to be completed using \thesystem. The tasks were related to the given scenario and the toy dataset and were designed to guide participants through a typical workflow of a data owner releasing statistics through the tool. We wrote tasks that required the effective use of each feature of the tool. Each task is listed below with the corresponding tested features in parentheses. 
\begin{enumerate}
\itemsep0em 
\footnotesize{
    \item You just entered a tutorial mode in the interface that will highlight some key features of the tool. Go through the tutorial and, when prompted, select a mean of the Age variable as your first statistic. (Tutorial mode, selecting statistics, setting metadata.)
    \item You decide that the income and race variables are also important for future researchers, so you decide to release statistics for these. Add a mean and a quantile for income, as well as a histogram for race. (Selecting statistics, setting metadata.)
	\item You no longer wish to include a quantile for income. Delete this statistic. (Deleting statistics.) 
	\item	You decide that you want to be very confident in your error estimates. Use the tool to set a 98 percent confidence level. (Adjusting confidence level.) 
	\item	You are thinking about your dataset, and you realize that it contains some information that makes it more sensitive than you originally thought. Use the tool to make the changes necessary to reflect this shift. (Adjusting privacy loss parameters, understanding that smaller privacy loss parameters gives greater privacy.) 
	\item	You have just been informed by a colleague that your dataset was actually randomly sampled from a population of size 1,200,000. Use the tool to make changes to reflect this. Does this make your statistics more or less accurate? (Secrecy of the sample.) 
	\item	You decide that it would be useful to allow other researchers who do not have access to your raw data to make some of their own selections for statistics to calculate from your dataset. Use the tool to make changes to reflect this. (Reserving privacy budget for data analysts.) 
	\item	How much error is there for the mean age statistic? What does this number mean? (Interpreting error estimates.)
	\item	Make it so that the released mean age is off from its true mean by at most one year. Is this more or less accurate than what you had before? (Redistributing privacy budget via error estimates.) 
	\item	Make it so that each count in the released race histogram is off from the true count by at most 5 people without changing the error you just set for mean age. (Redistributing privacy budget, hold feature.) 
	\item	You are satisfied with your statistics and your error estimates. Finalize your selections. (Submitting statistics.)
	}
\end{enumerate}

For each task we recorded three measures of performance: time spent on task, Critical Errors (CEs), and Non-Critical Errors (NCEs). Critical Errors are mistakes that resulted in the inability to complete the task at hand while Non-Critical Errors are mistakes that participants recovered from on their own and did not prevent the successful completion of a task. 

\subsection{Results}
\label{sect:results}
The results of the user test were encouraging. Participant performance on each task is summarized in Table \ref{tab:usability}. The most frequent mistakes occurred while entering metadata for statistics (Task 2). All participants expressed some doubt over how to enter metadata at some point during the study. 45\% of participants entered metadata values at some point during the study that would likely lead to an inaccurate statistic. 30\% of participants had trouble redistributing their privacy budget across the statistics they had selected. These results are discussed in more detail in Section \ref{sect:conveying}. Participants performed very well on the base functionalities of selecting and deleting statistics, modifying global parameters, and submitting the statistics for differentially private release. 

There was no significant relationship between the number of errors participants made (critical or non-critical) and familiarity with differential privacy or education level. These results suggest that \thesystem is usable for people of varied backgrounds as long as they have some experience with data analysis as all of our participants had. Unsolicited, four participants said that they had fun using the tool. Other results of the user study and lessons learned are discussed throughout Section \ref{sect:discussion}.

\begin{table}
\begin{center}
\begin{tabular}{|c|c|c|c|}
\hline
Task & Average time on task (secs)& \#CEs & \#NCEs \\
\hline
1 & 349.9 & 2 & 1  \\
\hline
2 & 289.7 &9  &6  \\
\hline
3 & 8.2 &2 &1  \\
\hline
4 &  29.8 & 2&  0\\
\hline
5 & 53.7 & 4 & 2  \\
\hline
6 &  30.1 &1 &0 \\
\hline
7 &  95.8 &5  &1 \\
\hline
8 & 34.3 & 2& 0  \\
\hline
9 &  20.3 &5  &3  \\
\hline
10 &  52.4 & 6 & 3  \\
\hline
11 & 20.6 &0  &0  \\
\hline
\end{tabular}
\end{center}
\caption{Performance on usability tasks. Average time on task is reported only for participants who successfully completed the task. The \#CEs and \#NCEs columns list the number of people who committed at least one critical error or non-critical error, respectively, during the task.}
\label{tab:usability}
\end{table}%

\subsection{System Usability Scale}
\label{sect:sus}
At the end of the study, each participant filled out the System Usability Scale (SUS) \cite{brooke96}, a questionnaire that is widely used in the UI testing community. The SUS has been shown to be a reliable and valid measure of usability across many types of interfaces \cite{sauro11, bkm08}. Comprised of ten five-point Likert scale questions, the SUS is easy to administer, minimizing participant fatigue. The SUS responses are combined into a single score between 0 and 100 that is not meant to be interpreted as a percentage or a letter grade. 

\begin{figure}[htb]
    \centering
   \includegraphics[width=7cm, height=7cm]{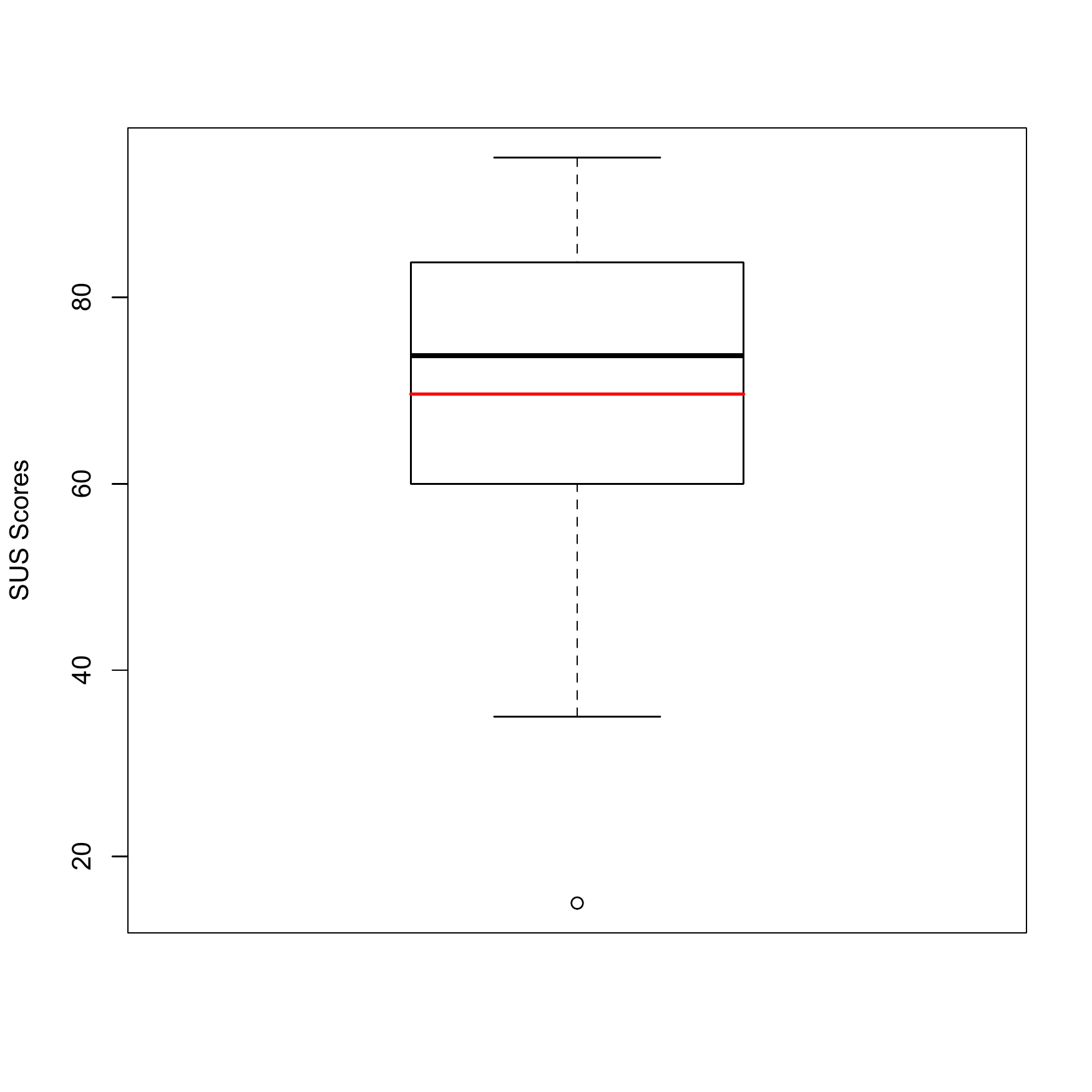}
    \caption{Box and whisker plot of System Usability Scale scores for \thesystem. The black horizontal line is the median score (73.6) and the red horizontal line is the mean score (69.6). The circle represents one outlier in the data.}
\label{fig:boxplot}
\end{figure}

The average SUS score given to \thesystem in the full user study was 69.6 (median = 73.6). Figure \ref{fig:boxplot} depicts the data in a box and whisker plot. One participant gave an outlying score more than 2.5 standard deviations below the mean. When this outlier is removed from the data, PSI's average SUS score jumps to 72.5 (median = 75.0). These scores mark a significant improvement from the pilot study, where PSI received an average score of 59.5. This suggests that improvements made to the tool between the pilot and full testing phases made PSI significantly easier to use. We found no significant relationship between participants' scores and their education level nor familiarity with differential privacy. For a breakdown of PSI's scores on the System Usability Scale by question, see Appendix \ref{app:sus}.

We can interpret these results based on past studies that used the SUS. Two meta-analyses have been conducted looking at hundreds of user studies that employed the SUS. They found that across all types of systems, the average score is 69.7 \cite{bkm08} and 68.0 \cite{sauro11}, respectively. Another group conducted a study that added an eleventh question to the SUS asking people to label the system with one of the following descriptors: Worst Imaginable, Awful, Poor, OK, Good, Excellent, Best Imaginable \cite{bkm08}. They found that of the group that selected OK, the mean assigned SUS score was 52.0, while the average score of the Good systems was 72.8. Some well known user interfaces have been evaluated using the SUS, including Excel, which received a mean score of 56.5, GPS systems, which received 70.8, and the iPhone, which scored 78.5 on average \cite{kb13}.  Given that \thesystem aims to solve a complex problem, we find the results on the SUS encouraging.

\subsection{Improving PSI}
Between each phase of user tests, we made adjustments to PSI's interface in response to participant feedback and performance. One large change was the introduction of a tutorial mode that is automatically triggered upon first entering the tool. The tutorial guides users around the features of the page and parcels out the documentation in small digestible blocks.

We also made significant modifications to the documentation itself, emphasizing clear and concise language and rewriting specific points of confusion observed during the tests. There were also several additional smaller changes to make the tool more user-friendly such as new buttons, bug fixes, hiding advanced features when they're not needed, and a host of cosmetic adjustments. 

We paid special attention to changes intended to prevent privacy leaks. Most significantly, we implemented rigorous and automatic error-checks at the point of privacy loss parameter selection to warn data owners when they make unusual or unsafe choices.

\section{Discussion}
\label{sect:discussion}
The purpose of this paper is to communicate the insights we gained from conducting a usability study of \thesystem in hopes that they will inform future efforts to implement differential privacy in user-facing settings, as well as contribute to the larger body of knowledge of privacy practitioners and system designers. In this section, we discuss the lessons learned from our study, as well as the challenges we faced and general insights regarding bringing differential privacy to practice. 

\subsection{Conveying Differential Privacy to Users}
\label{sect:conveying}
In designing and implementing the usability study of \thesystem, we identified several areas specific to differential privacy that are challenging to convey to lay users. These are likely to be relevant for future implementations of differential privacy.
\subsubsection{Privacy Loss Parameters}
Setting the privacy loss parameters ($\epsilon$ and $\delta$) is the first and arguably most important task in the \thesystem interface because they govern all subsequent calculations and releases, and can cause the greatest harm to privacy. 

\thesystem provides users with detailed explanations and guidance on the privacy loss parameters including a table interpreting the parameters in terms of changes in posterior belief about the likelihood that a subject has a given trait. 19 out of 20 participants in the formal usability study successfully entered privacy loss parameters when prompted, choosing appropriate values according to the tool's recommendations (see Section \ref{sect:safeguards} for a description of the 1 participant who mis-entered the privacy loss parameters). 

Despite this, there were some mistakes on subsequent tasks that suggested that a small group of users had not digested the relationship between the privacy loss parameters and the sensitivity-level of their dataset. One of the tasks told participants, ``You are thinking about your dataset, and you realize that it contains some information that makes it more sensitive than you originally thought. Use the tool to make the changes necessary to reflect this shift'', with the goal of guiding them to edit the privacy loss parameters to make them smaller, and therefore more protective of privacy. 4 out of 20 participants in the formal usability study committed one or more critical errors on this task, which in this case involved attempting to use a different feature of the tool to manipulate privacy, such as changing the confidence level or editing the individual error estimates.

This pattern of errors only appeared in one fifth of our tests, and did not ultimately result in significant privacy loss. We nevertheless highlight these errors related to the privacy loss parameters because they are the most impactful errors that can be made in a differential privacy system, and should be minimized as much as possible through clear explanations, intuitive design features, and rigorous error checking. 

The first PSI prototype included the privacy loss parameters as editable text on the main screen of the interface. We soon realized, however, that this feature could tempt data owners to dangerously increase the privacy loss parameters if, say, they were dissatisfied with the error estimates for their statistics. To mitigate this risk, we moved the editable parameters from the main screen to a separate window that could be accessed by clicking a somewhat inconspicuous button at the corner of the screen titled ``Edit Parameters.'' Such an adjustment conveys that the overall privacy loss parameters should be considered more of a fixed element in the calculations, rather than something that users should edit while making their selections for statistics and error levels.

\subsubsection{Metadata}
As discussed in Section \ref{sect:results}, entering metadata was the most common source of confusion and error during our usability test. Our analysis revealed that 9 out 20 participants entered metadata values that would likely lead to inaccurate statistics. Almost all of these errors were easily preventable by clearer documentation and examples, which have since been incorporated into the tool. Other systems like \thesystem should be conscientious of this common stumbling block and should provide explanations and recommendations adequate to prevent metadata errors that could lead to accuracy losses.  

One potential way to mitigate this issue is through the use of codebooks. Many datasets in the social sciences are accompanied by codebooks that list the possible values for each variable in the dataset. For example, multiple choice surveys that are conducted repeatedly may use codebooks to make their data interpretable to outsiders or to standardize responses across research groups. If there were a tool that could parse such codebooks and use them to pre-populate metadata fields in a differential privacy system, it would greatly reduce user fatigue and the likelihood of user error. We think this is an interesting research direction in this space. 

The handling of metadata is also a non-obvious area where users can accidentally cause privacy loss by using values directly from their raw data. 
To prevent this, the \thesystem interface tells users in bold font not to refer directly to their raw dataset, but rather to fill in general, a priori estimates. This text can be seen in the interface in Figure \ref{fig:interface}.

\subsubsection{Budgeting Privacy and Accuracy}
Arguably the most complex aspect of bringing differential privacy into practice within a user interface is conveying the ideas of a tradeoff between privacy and accuracy, the limiting of privacy and accuracy with a budget, the distribution of that budget across multiple statistics, and the reserving of portions of the budget for future use by data analysts. However full internalization of these relationships may not be essential for the successful use of \thesystem or similar systems. 

The primary goal of \thesystem is not to teach people about the intricacies of differential privacy, but rather to help them use it effectively. So the interface and documentation aims to teach users concepts insofar as it will help them achieve their ultimate goals of releasing accurate statistics while maintaining privacy. This distinction became clear in the user tests especially around the concept of a privacy budget. In total, 11 of the 20 participants in the formal usability study explicitly voiced some confusion about the concept of a ``budget'' for privacy and accuracy. However these 11 participants still completed an average of 8.9 of our 11 tasks correctly (without critical error). This suggests that people without a complete understanding of the more high-level concepts of differential privacy are still able to work with it successfully. In efforts seeking to bring differential privacy to a wider audience, consideration should be given to how much material should be attempted to be explained, and how many features or controls should be presented to the user. Fewer moving parts might lessen confusion about the relationships between privacy, accuracy, and budgeting.

\subsection{Safeguards Against User Errors}
\label{sect:safeguards}
Many use cases of differential privacy, such as those discussed in Section \ref{sect:systems}, do not involve putting privacy-critical decisions into the hands of lay people. As differential privacy continues to expand, however, we can expect more use cases like \thesystem that require important input by non-experts. Such settings necessitate precise documentation, intuitive design, and thorough error-handling.

There are three ways within \thesystem in which data owners could inadvertently compromise the privacy of their data subjects:
\begin{enumerate}
\item Overestimating the population size in the secrecy of the sample feature.
\item Entering empirical values from the raw data for metadata.
\item Setting inappropriately large privacy loss parameters.
\end{enumerate}
The secrecy of the sample feature is discussed more in Section \ref{sect:secofsamp}. Not one of the 28 participants across the 3 study phases overestimated the population size in using the secrecy of the sample feature even though a majority of the participants elected to use it. 

One pilot participant used empirical values from the raw data when entering metadata for a statistic. In response to this, we added additional text in the interface at the point of metadata elicitation, warning data owners about the risks of providing data-dependent values (seen in Figure \ref{fig:interface}). After adding this text, all 20 participants in the full study resisted referring to the raw data, even when verbally expressing a temptation to do so.

Setting overly large privacy loss parameters is a concern for all differential privacy systems. In our user tests, only one of the 28 participants chose unwise privacy loss parameters by accidentally switching the intended values for the two parameters (setting $\epsilon=10^{-6}$ and $\delta=.25$). Every other participant followed the recommendations in the documentation, setting $\epsilon$ between $.05$ and $1$ and $\delta$ between $10^{-7}$ and $10^{-5}$. Although the mistake was a rare event, we took it seriously and implemented more rigorous error checks at the point of parameter selection, warning users when their parameters seem unsafe.

These results suggest that \thesystem is designed in such a way that makes it difficult to violate privacy even for users who might make many mistakes while using the tool.  In other words, even when users made choices that could lead to poor accuracy in the outputs, it was very rare that users made choices that could compromise privacy. In many cases it is important that a differentially private system prioritizes privacy above accuracy. The goal should be to design a system in which it is difficult to violate privacy even when misused.

\subsection{Suggestibility of Users}
Selection of privacy loss parameters is a critical decision in any differential privacy system. For many applications, this decision can be made by experts taking into account the type, size, and sensitivity level of the data being protected. PSI puts this decision into the hands of its users, who are not expected to be trained in privacy, computer science, or statistics. As a guide, the version of the tool used in the study describes different levels of data sensitivity inspired by Harvard University's 5 Data Security Levels\footnote{https://policy.security.harvard.edu/view-data-security-level} and assigns recommended settings of privacy loss parameters for each as follows:

%\itemsep0em 
%\footnotesize{
\begin{enumerate}
\item Public information: It is not necessary to use differential privacy for public information.
\item Information the disclosure of which would not cause material harm, but which the University has chosen to keep confidential: ($\epsilon=1,~\delta=10^{-5}$)
\item Information that could cause risk of material harm to individuals or the University if disclosed: ($\epsilon=.25,~\delta=10^{-6}$) 
\item Information that would likely cause serious harm to individuals or the University if disclosed: ($\epsilon=.05,~\delta=10^{-7}$)
\item Information that would cause severe harm to individuals or the University if disclosed: It is not recommended that the PSI tool be used with such severely sensitive data.
\end{enumerate}
%}
Of the 20 participants in the full study, 16 of them chose their privacy loss parameters from the three options above. 1 out of the 20 accidentally set $\epsilon=10^{-6}$ and $\delta=.25$, as discussed in Section \ref{sect:safeguards} above, and only 3 out of the 20 chose numbers that were not listed above (but remained within the suggested ranges). 

These results highlight the importance of parameter suggestions in differential privacy tools. Untrained data owners are likely to set parameters exactly as recommended rather than determine settings appropriate for their dataset based on technical definitions. The recommendations in PSI are based on the experience of the developers and aim to be conservative. However, choosing the right privacy loss parameters for a given set of statistics is a non-trivial task and improvements in guiding lay users through the selection process is an important research direction as differential privacy becomes more mainstream.

\subsection{Secrecy of the Sample}
\label{sect:secofsamp}
As stated in Lemma \ref{lem:secofsamp}, Secrecy of the Sample is a technique that boosts the privacy guarantee of an algorithm by first randomly sampling a subset of records from the input dataset and then running a differentially private computation on the sample only. This boosts privacy because each record in the data has a non-negligible probability of contributing zero information to the output of the algorithm because it may not be in the sample. 

In science, datasets usually contain a random subset of some larger population. So a useful interpretation of secrecy of the sample is to view the population as the larger dataset and the actual input to a differentially private algorithm as a subsample of the population. As long as this subsample is truly random and the choice of the people in the subsample remain secret, the privacy amplification from secrecy of the sample comes for free. Because of the privacy/accuracy trade-off, these savings can actually be viewed as free boosts in accuracy while maintaining the same level of privacy.  

This technique is incorporated into PSI to provide accuracy boosts for data owners whose datasets fit the criteria for secrecy of the sample. The prompt given to participants in the user study contained the population size and the fact that the data were randomly sampled from the population. The secrecy of the sample feature was described in the documentation of the tool but was left optional and was never explicitly mentioned by us during the study. Of the 20 participants, 13 opted to use the feature because they knew the relevant criteria and perceived it to improve the accuracy of their statistics. 

Secrecy of the sample is a powerful technique that can mitigate one of the challenges to bringing differential privacy to practice: outputs that are too noisy. However it must be employed with care. If the dataset is not a truly IID random sample, the population size is overestimated, or the membership of an individual in the sample dataset can leak, the conditions for secrecy of the sample are violated and it should not be used. 

\subsection{Relevance of Privacy}
Another barrier to bringing differential privacy to practice is convincing people that there is a problem to be solved in the first place. Only then can one hope to motivate people to put in the extra work needed to use a tool like \thesystem. At the end of the usability test, participants were asked how relevant they think \thesystem is for researchers with privacy sensitive datasets on a scale from 1 to 5. The average response given was 4.3. We view this as a sign that people who work with human-subjects data by and large are aware of privacy concerns in their fields, recognize the inadequacy of currently available solutions (e.g. removing ``personally identifiable information'' (PII)), and see \thesystem as a system that addresses a need in this space.

\subsection{Designing Usability Tests for Differential Privacy Systems}
Our study yielded several insights related to the design of usability tests in the context of differential privacy. The first is generated by the question of how to frame tasks within the test. The initial version of our protocols, used in the pilot phase, presented the participant with an open-ended scenario. Specifically, we provided the participant with a dataset and told them to behave as if they were a data owner who had collected the data and wished to release some statistics about them while protecting privacy. Then, we asked the participant to use \thesystem to release differentially private statistics about their dataset. We did not provide further instructions.

Participants in the pilot test seemed overwhelmed, and we realized that much of this came from the lack of structure in our test protocols. These participants were learning about differential privacy for the first time, and then immediately being asked to work with it in an abstract scenario. This set-up did not accurately simulate the experience of a real user of \thesystem, because an actual data owner would already have a clear vision for their goals working with the tool, such as the level of sensitivity of the data and the most important variables and statistics. Our loose instructions did not provide participants with a similarly clear mindset that would allow them to progress through the interface with ease. Our protocols thus needed to evolve to include sequential tasks to guide participants' intentions and lead them through each feature in the interface.

The open-endedness of our protocols at the pilot stage also made it difficult to systematically collect numerical data to represent the performance of our pilot participants. For the formal study, we used clear, specific tasks that would lead to quantifiable outcomes corresponding to each component of \thesystem.

These insights informed our subsequent revision of the test protocols to produce the final version that we employed with 20 participants in the formal usability study. The updated protocols eliminated the single prompt of the pilot test which asked participants to use the interface to release some statistics about their dataset. Instead, we devised the 11 tasks listed in Section \ref{sect:protocol}, each tailored to assess the experience of the participants with specific components of the interface. We timed the participants on each task, while observing them in order to note all critical and non-critical errors. We derived our definitions of critical and non-critical errors from usability.gov.\footnote{https://www.usability.gov/how-to-and-tools/methods/planning-usability-testing.html} 

These new protocols allowed us to quantify the success of our study participants. However, differential privacy complicates the notion of success in usability studies, as there are several areas in which it is difficult to classify user behavior as ``correct'' or ``incorrect''. Specifically, although the interface presents guidelines for setting the privacy loss parameters, there is no universally accepted standard. For example, the highest (least private) recommended value for $\epsilon$ in \thesystem is 1, while it was recently revealed that Apple’s implementation of differential privacy employed $\epsilon$ values as high as 14.\footnote{https://www.wired.com/story/apple-differential-privacy-shortcomings/} Additionally, there are many components of differential privacy in practice that exist on a spectrum of choice, making it difficult to determine when exactly a user has done something ``wrong''. These include choosing which statistics to release, determining how to distribute error across those statistics, and deciding how much of the global privacy budget to exhaust at a given time. These gray areas necessitate even more rigorous operationalization of successes and errors on tasks in usability studies involving differential privacy.

Still a relatively new concept, differential privacy presents a multitude of questions in terms of its implementation. Current and future evaluations of the utility of differentially private algorithms, as well as explorations of the range of appropriate applications for real-life data analysis, are necessary and important. It is a separate endeavor to study the usability of the interface used to present differential privacy, but it can be difficult not to conflate this goal with the others. Moreover, it is likely that the underlying concepts of differential privacy are entirely foreign to the participants, and hence, it is difficult in a usability study to distinguish errors due to confusion about differential privacy from errors due to a poor interface. Each of these types of errors matter but they may have different solutions. To mitigate this, we made a special effort to delineate clear tasks corresponding to specific features of the interface, and to urge participants to vocalize their thoughts as much as possible, asking clarifying questions where necessary. 

\section{Conclusion and Open Questions}
We conducted a usability study of \thesystem, an interface that aims to bring differential privacy into wide use by allowing data owners to calculate privacy-preserving statistics about their datasets. The study shed light on some of the challenges in making differential privacy accessible to non-experts, on the quality of PSI's particular attempt at such implementation, and on the techniques for designing and conducting informative usability tests for differential privacy.

Our findings highlight the suggestibility of users, the importance of carefully explaining the purpose and significance of the privacy loss parameters, the likelihood of misunderstanding of metadata requirements, and the need for clear, specific test protocols to allow for quantitative results and to make differential privacy less intimidating to test participants. We hope that these lessons will be of use to others who aim to deploy differential privacy in user-facing settings.

\subsection{Open Questions}

Though the \thesystem system attempts to address this as much as possible, the question remains of how best to explain information and present options related to the highly technical aspects of differential privacy. There is a trade-off between the number of technical choices a system provides to users and the ability of those users to operate the system without confusion or error. The more choices available, the more complex the environment becomes and the more explanatory text is necessary to orient users. This can render an interface bulky and intimidating. On the other hand, the full benefits of differential privacy may only be realized if people are able to apply it agilely and in a variety of contexts, requiring a tool with flexibility and a variety of options.

Another consideration for future work is that, though they are of the utmost importance, privacy loss parameters are not currently standardized in terms of the numerical values that should correspond to each level of data sensitivity. Because errors relating to the privacy loss parameters have the greatest risk of harming privacy, explanations must be firm and straightforward, and guidance should be provided wherever possible. However, as discussed in this paper, there exists wide discrepancy in the setting of privacy loss parameters in existing deployments of differential privacy. It is difficult to simultaneously convey the care with which these parameters must be handled and the lack of consensus on the appropriate values to date.

Finally, we acknowledge that \thesystem's features capture only a fraction of the potentially vast range of calculations and applications available for differential privacy, and we urge future efforts to creatively explore new ways to bring differential privacy into practice. \\

\paragraph{Acknowledgements} We would like to thank Derek Murphy and Tania Schlatter for very useful advice and feedback at the early stages of this work. 

\newpage

\bibliographystyle{abbrv}
\bibliography{sigproc}

\begin{thebibliography}{10}

\bibitem{bkm08}
A.~Bangor, P.~Kortum, and J.~T. Miller.
\newblock An empirical evaluation of the system usability scale.
\newblock {\em International Journal of Human--Computer Interaction},
  24(6):574--594, 2008.

\bibitem{brooke96}
J.~Brooke.
\newblock Sus-a quick and dirty usability scale.
\newblock {\em Usability evaluation in industry}, 189(194):4--7, 1996.

\bibitem{Crosas11}
M.~Crosas.
\newblock The dataverse network: An open-source application for sharing,
  discovering and preserving data.
\newblock {\em D-Lib Magazine}, 17(1), 2011.

\bibitem{DworkKeMcMiNa06}
C.~Dwork, K.~Kenthapadi, F.~McSherry, I.~Mironov, and M.~Naor.
\newblock Our data, ourselves: Privacy via distributed noise generation.
\newblock In {\em EUROCRYPT}, pages 486--503, 2006.

\bibitem{dwork06}
C.~Dwork, F.~McSherry, K.~Nissim, and A.~Smith.
\newblock Calibrating noise to sensitivity in private data analysis.
\newblock In {\em TCC}, 2006.

\bibitem{DRV10}
C.~Dwork, G.~N. Rothblum, and S.~P. Vadhan.
\newblock Boosting and differential privacy.
\newblock In {\em FOCS}, pages 51--60, 2010.

\bibitem{erlingsson14}
{\'U}.~Erlingsson, V.~Pihur, and A.~Korolova.
\newblock Rappor: Randomized aggregatable privacy-preserving ordinal response.
\newblock In {\em Proceedings of the 2014 ACM SIGSAC conference on computer and
  communications security}, pages 1054--1067. ACM, 2014.

\bibitem{PSI}
M.~Gaboardi, J.~Honaker, G.~King, K.~Nissim, J.~Ullman, and S.~P. Vadhan.
\newblock {PSI} ({\(\Psi\)}): a private data sharing interface.
\newblock {\em CoRR}, abs/1609.04340, 2016.

\bibitem{haeberlen11}
A.~Haeberlen, B.~C. Pierce, and A.~Narayan.
\newblock Differential privacy under fire.
\newblock In {\em USENIX Security Symposium}, 2011.

\bibitem{honaker2014statistical}
J.~Honaker and V.~D'Orazio.
\newblock Statistical modeling by gesture: A graphical, browser-based
  statistical interface for data repositories.
\newblock In {\em Extended Proceedings of ACM Hypertext 2014}. ACM, 2014.

\bibitem{KLNRS08}
S.~P. Kasiviswanathan, H.~K. Lee, K.~Nissim, S.~Raskhodnikova, and A.~Smith.
\newblock What can we learn privately?
\newblock {\em SIAM J. Comput.}, 40(3):793--826, 2011.

\bibitem{King07}
G.~King.
\newblock An introduction to the dataverse network as an infrastructure for
  data sharing.
\newblock {\em Sociological Methods and Research}, 36:173--199, 2007.

\bibitem{kb13}
P.~Kortum and A.~Bangor.
\newblock Usability ratings for everyday products measured with the system
  usability scale.
\newblock {\em International Journal of Human-Computer Interaction},
  29(2):67--76, 2013.

\bibitem{mac08}
A.~Machanavajjhala, D.~Kifer, J.~Abowd, J.~Gehrke, and L.~Vilhuber.
\newblock Privacy: Theory meets practice on the map.
\newblock In {\em Data Engineering, 2008. ICDE 2008. IEEE 24th International
  Conference on}, pages 277--286. IEEE, 2008.

\bibitem{mcsherry09}
F.~D. McSherry.
\newblock Privacy integrated queries: an extensible platform for
  privacy-preserving data analysis.
\newblock In {\em Proceedings of the 2009 ACM SIGMOD International Conference
  on Management of data}, pages 19--30. ACM, 2009.

\bibitem{mohan12}
P.~Mohan, A.~Thakurta, E.~Shi, D.~Song, and D.~Culler.
\newblock Gupt: privacy preserving data analysis made easy.
\newblock In {\em Proceedings of the 2012 ACM SIGMOD International Conference
  on Management of Data}, pages 349--360. ACM, 2012.

\bibitem{MurtaghV16}
J.~Murtagh and S.~P. Vadhan.
\newblock The complexity of computing the optimal composition of differential
  privacy.
\newblock In {\em Theory of Cryptography - 13th International Conference, {TCC}
  2016-A, Tel Aviv, Israel, January 10-13, 2016, Proceedings, Part {I}}, pages
  157--175, 2016.

\bibitem{roy10}
I.~Roy, S.~T. Setty, A.~Kilzer, V.~Shmatikov, and E.~Witchel.
\newblock Airavat: Security and privacy for mapreduce.
\newblock In {\em NSDI}, volume~10, pages 297--312, 2010.

\bibitem{sauro11}
J.~Sauro.
\newblock {\em A practical guide to the system usability scale: Background,
  benchmarks \& best practices}.
\newblock Measuring Usability LLC, 2011.

\bibitem{AdamSecrecyPost}
A.~Smith.
\newblock Differential privacy and the secrecy of the sample.
\newblock \url{https://adamdsmith.wordpress.com/2009/09/02/sample-secrecy/}.

\bibitem{apple}
J.~Tang, A.~Korolova, X.~Bai, X.~Wang, and X.~Wang.
\newblock Privacy loss in apple's implementation of differential privacy on
  macos 10.12.
\newblock {\em CoRR}, abs/1709.02753, 2017.

\end{thebibliography}

\appendix
\section{System Usability Scale}\label{app:sus}
Table \ref{tab:sus} shows the average score \thesystem received on each question in the SUS questionnaire, discussed in Section \ref{sect:sus}. The questionnaire was given to all participants at the end of the study. There are 10 items with five numerical response options for respondents, ranging from Strongly Disagree (1) to Strongly Agree (5).

\begin{table*}[!ht]
\begin{center}
\begin{tabular}{|r|l|c|}
\hline
 &\bf{SUS Item} & \bf{Avg. score} \\
\hline
\hline
1&I think that I would like to use this system frequently. & 3.75  \\
\hline
2&I found the system unnecessarily complex. & 2.15 \\
\hline
3&I thought the system was easy to use. & 3.6 \\
\hline
4&I think that I would need the support of a technical person to be able to use this system. &  2.35\\
\hline
5 &I found the various functions in this system were well integrated.&  4.2 \\
\hline
6&I thought there was too much inconsistency in this system.
 &  1.8 \\
\hline
7&I would imagine that most people would learn to use this system very quickly. &  3.6 \\
\hline
8&I found the system very cumbersome to use. & 1.6 \\
\hline
9&I felt very confident using the system. &  3.3  \\
\hline
10&I needed to learn a lot of things before I could get going with this system. &  2.7  \\
\hline
\end{tabular}
\end{center}
\caption{Average System Usability Scale (SUS) scores for \thesystem broken down by question.}
\label{tab:sus}
\end{table*}%
\end{document}